\documentclass[%
 reprint,
nofootinbib,
 amsmath,amssymb,
 aps,
]{revtex4-1}

\usepackage{graphicx}% Include figure files
\usepackage{dcolumn}% Align table columns on decimal point
\def\hyp{\mathsf{y}}
\usepackage{bm}% bold math
\usepackage[multiple]{footmisc}
\newcommand{\Lagr}{\mathcal{L}}

\begin{document}

%\preprint{APS/123-QED}
\title{Gauge fixing the Standard Model Effective Field Theory}% Force line breaks with \\

\author{Andreas Helset$^a$}
\email{ahelset@nbi.ku.dk}
\author{Michael Paraskevas$^b$}
\email{mparask@grads.uoi.gr}
\author{Michael Trott$^a$}
\email{michael.trott@cern.ch}
\affiliation{%
a) Niels Bohr International Academy and Discovery Centre, Niels Bohr Institute,
 University of Copenhagen, Blegdamsvej 17, DK-2100 Copenhagen, Denmark \\
b) Department of Physics, University of Ioannina,
GR 45110, Ioannina, Greece
}%

\date{\today}% It is always \today, today,

\begin{abstract}
We gauge fix the Standard Model Effective Field Theory in a manner invariant under background field gauge transformations
using a geometric description of the field connections.
%\begin{description}
%\end{description}
\end{abstract}

\maketitle

%\tableofcontents

{\bf Introduction.}
When physics beyond the Standard Model (SM) is present at scales ($\Lambda$) larger than the Electroweak scale
($\sqrt{2 \, \langle H^\dagger H} \rangle \equiv \bar{v}_T$), the SM can be extended into an effective field theory (EFT).
The Standard Model Effective Field Theory (SMEFT), defined by a power counting expansion
in the ratio of scales $\bar{v}_T/\Lambda <1$, extends the SM with higher dimensional operators $\mathcal{Q}_i^{(d)}$
of mass dimension $d$. The Lagrangian is
\begin{align}
	\Lagr_{\textrm{SMEFT}} &= \Lagr_{\textrm{SM}} + \Lagr^{(5)}+\Lagr^{(6)} +
	\Lagr^{(7)} + \dots,  \\ \nonumber \Lagr^{(d)} &= \sum_i \frac{C_i^{(d)}}{\Lambda^{d-4}}\mathcal{Q}_i^{(d)}
	\textrm{ for } d>4.
\end{align}
The SMEFT is a model independent and consistent low energy parameterization of heavy physics beyond the SM, so long as its defining assumptions are satisfied:
that there are no light hidden states in the spectrum with couplings
to the SM; and a $\rm SU(2)_L$ scalar doublet with hypercharge
$\hyp_h = 1/2$ is present in the EFT.

The SMEFT has the same $\rm SU(3)_C \times SU(2)_L \times U(1)_Y$ global symmetry as the SM.
The SMEFT also has a Higgsed phase of $\rm SU(2)_L \times U(1)_Y \rightarrow U(1)_{em}$. A difference between these theories
is that additional couplings and interactions between the fields
come about due to the $\mathcal{Q}_i^{(d)}$.
Some of these interactions are bilinear in the SM fields in the Higgsed phase.
These terms are important for gauge fixing and the presence of these interactions introduce technical
challenges to the usual gauge fixing approach.

The bilinear field interactions in the SMEFT are usefully thought of in
terms of connections in the field space manifolds of the theory \cite{Alonso:2015fsp,Alonso:2016oah}.
The purpose of this paper is to show that gauge fixing the SMEFT, taking into account the field space
metrics, directly resolves many of the technical challenges that have been identified to date.
The approach we develop generalizes directly to higher orders in the SMEFT power counting expansion.

The difficulties in gauge fixing the SMEFT are also present when the Background Field Method (BFM)
\cite{DeWitt:1967ub,tHooft:1973bhk,Abbott:1981ke,Shore:1981mj,Einhorn:1988tc,
Denner:1994nn,Denner:1994xt} is used \cite{Hartmann:2015oia}. The BFM splits the fields in the theory into
quantum and classical fields ($F \rightarrow F + \hat{F}$),
with the latter denoted with a hat superscript. One performs a gauge fixing procedure that preserves background field gauge invariance while breaking
explicitly the quantum field gauge invariance. This allows a gauge choice for the quantum fields to be made to one's advantage, while
still benefiting from the simplifications that result from naive Ward identities \cite{Ward:1950xp} due to the preserved background field gauge
invariance.\footnote{The Ward identities result from considering BRST invariance \cite{Becchi:1975nq} when the BFM is not used, which can be more cumbersome when extending results to higher orders
in the SMEFT power counting expansion.}

In this paper, we show how to perform gauge fixing with the BFM taking into account the field space metrics that are present due to
the SMEFT power counting expansion.
The usual $R_\xi$ gauge fixing approach in the BFM for the Standard Model \cite{Shore:1981mj,Einhorn:1988tc,Denner:1994nn,Denner:1994xt}
is a special case of this approach.\footnote{For a $R_\xi$
gauge SMEFT formulation with three distinct $\xi$ parameters see Ref.~\cite{Dedes:2017zog}.} Conceptually one can understand that this procedure is advantageous as it preserves
the background
$\rm SU(3)_C \times SU(2)_L \times U(1)_Y$ invariance on the curved field spaces present due to the power counting expansion.
The latter is trivialized away in the Standard Model.

{\bf Scalar space.}
The operators that lead to scalar kinetic terms in the Higgsed phase of the theory up to $\mathcal{L}^{(6)}$ are \cite{Grzadkowski:2010es}
\begin{align}\label{scalarL6}
	\Lagr_{\textrm{scalar,kin}} =& \left(D_{\mu} H\right)^\dagger \left(D^{\mu}H\right)
	+ \frac{C_{H\Box}}{\Lambda^2}\left(H^\dagger H\right)
	\Box \left(H^\dagger H\right) \nonumber \\
	&+ \frac{C_{HD}}{\Lambda^2}\left(H^\dagger D_{\mu}H\right)^*
	\left(H^{\dagger}D^{\mu} H\right), \nonumber \\
	\equiv& \frac{1}{2}h_{IJ}(\phi)\left(D_{\mu}\phi\right)^I\left(D^{\mu}\phi\right)^J.
\end{align}
Our covariant
derivative sign convention is given by $D^\mu H = (\partial^\mu + i \, g_2 W^{a, \mu} \,\sigma_a/2 + i \,g_1 \, \hyp_h B^\mu Y)H$ and 
$(D^{\mu}\phi)^I = (\partial^{\mu}\delta_J^I - \frac{1}{2}W^{A,\mu}\tilde\gamma_{A,J}^I)\phi^J$, with
definitions given below.
Defining
\begin{align}
	H = \frac{1}{\sqrt{2}}\begin{bmatrix} \phi_2+i\phi_1 \\ \phi_4 - i\phi_3\end{bmatrix},
\end{align}
the scalar field connections can be described by a $\mathbb{R}^4$ field manifold with the metric $h_{IJ}(\phi)$.
Our notation is that the latin capital letters  $I,J,K,L \cdots$ run over $\{1,2,3,4\}$, while lower
case latin letters $i,j,k,l \cdots$ run over $\{1,2\}$.
The metric takes
the form
\begin{align}
	h_{IJ}(\phi) = \delta_{IJ} - 2\frac{C_{H\Box}}{\Lambda^2}\phi_I\phi_J
	+ \frac{1}{2}\frac{C_{HD}}{\Lambda^2}f_{IJ}(\phi),
\end{align}
where
\begin{align}
	f_{IJ}(\phi) =& \begin{bmatrix}
		a & 0 & d & c \\
		0 & a & c & -d \\
		d & c & b & 0 \\
		c & -d & 0 & b
	\end{bmatrix}, \quad
	\begin{aligned}
		a &= \phi_1^2 + \phi_2^2, \\
		b &= \phi_3^2 + \phi_4^2,\\
		c &= \phi_1\phi_4 + \phi_2\phi_3,\\
		d &= \phi_1\phi_3 - \phi_2\phi_4.
	\end{aligned}\\ \nonumber
  \end{align}
The Riemann curvature tensor calculated
from the scalar field metric is
non-vanishing  \cite{Burgess:2010zq,Alonso:2015fsp,Alonso:2016oah}. The scalar manifold is curved due to the power counting expansion.
An interesting consequence is that there does not exist
a gauge independent field redefinition which sets $h_{IJ} = \delta_{IJ}$  when considering $\mathcal{L}^{(6)}$
corrections \cite{Burgess:2010zq}.
As a result, demanding that the Higgs doublet field to be canonically normalized in the SMEFT to $\mathcal{L}^{(6)}$
cannot be used as a defining condition for operator bases \cite{Burgess:2010zq,Passarino:2016pzb,Passarino:2016saj,Brivio:2017vri}.

%We can also express the scalar kinetic term in terms of the $H$ doublet,
%
%\begin{align}
%	\Lagr_{\textrm{scalar,kin}} = h_{ij}(D_{\mu}H^{\dagger})^i(D^{\mu}H)^j, \\
%	h_{ij} = \begin{bmatrix}
%		1+a & -c+id \\
%		-c + id & 1+b
%	\end{bmatrix}
%\end{align}
%where
%\begin{align}
%	a &= \frac{1}{2}\frac{C_{HD}}{\Lambda^2}(\phi_1^2 + \phi_2^2) \nonumber \\
%	b &= \frac{1}{2}\frac{C_{HD}}{\Lambda^2}(\phi_3^2 + \phi_4^2) \nonumber \\
%	c &= \frac{1}{2}\frac{C_{HD}}{\Lambda^2}(\phi_2\phi_3 + \phi_1\phi_4) \nonumber \\
%	d &= \frac{1}{2}\frac{C_{HD}}{\Lambda^2}(\phi_1\phi_3 - \phi_2\phi_4).
%\end{align}
{\bf Gauge boson space.}
The operators that lead to $\rm CP$ even bilinear interactions for the $\rm SU(2)_L \times U(1)_Y$ spin one fields up to $\mathcal{L}^{(6)}$
are
\begin{align}\label{WBlagrangian}
	\Lagr_{\textrm{WB}} &= -\frac{1}{4}W_{\mu\nu}^aW^{a,\mu\nu}-\frac{1}{4}B_{\mu\nu}B^{\mu\nu}
	+  \frac{C_{HB}}{\Lambda^2}H^\dagger HB_{\mu\nu}B^{\mu\nu} \nonumber\\
	&+ \frac{C_{HW}}{\Lambda^2}H^\dagger HW_{\mu\nu}^aW^{a,\mu\nu}
	+ \frac{C_{HWB}}{\Lambda^2}H^\dagger \sigma^aHW_{\mu\nu}^aB^{\mu\nu}, \nonumber \\
	&\equiv-\frac{1}{4}g_{AB}(H) \mathcal{W}_{\mu\nu}^A \mathcal{W}^{B,\mu\nu},
\end{align}
where $a,b \cdots$ run over $\{1,2,3\}$, $A,B,C \cdots$ run over $\{1,2,3,4\}$.
Here $\mathcal{W}_{\mu\nu}^4=B_{\mu\nu}$.
Analogous to the scalar sector,
we have introduced a metric
$g_{AB}[H(\phi_i)]$, taking the form

\begin{align}
	g_{ab} &= \left(1-4\frac{C_{HW}}{\Lambda^2}H^\dagger H\right)\delta_{ab}, & g_{44} &= 1-4\frac{C_{HB}}{\Lambda^2}H^\dagger H, \nonumber \\
	g_{a4} &= g_{4a}  =  -2\frac{C_{HWB}}{\Lambda^2}H^\dagger\sigma_a H.
\end{align}
The Riemann curvature tensor
for the gauge fields can be calculated from $g_{AB}$ and is nonvanishing; the ($\rm CP$ even) $\mathbb{R}^4$ spin one field manifold
is also curved.\footnote{$\rm SU(2)_L$ is self adjoint. As a result, one can define
a $\mathbb{G}^{AB}$ tensor of the same form as $g_{AB}$ through
$\mathbb{G}^{AB}(H) \, \mathcal{W}^{\mu\nu}_A \, \mathcal{W}_{B,\mu\nu}$. This $\mathbb{G}^{AB}$ is not the
tensor $g^{AB}$ defined through the relation $g^{AB} g_{BC} =  \delta^A_C$ and used in the gauge fixing term.} A physical consequence is that, as in the case of the scalar manifold, there does not
exist a gauge independent field redefinition that sets $g_{AB} = \delta_{AB}$ including $\mathcal{L}^{(6)}$ corrections.\footnote{
A rotation to the mass eigenstate basis for the field bilinear interactions can be made,
and this is consistent with the curvature of the gauge manifold.}$^,$\footnote{Field redefiniton invariant quantities are more directly connected to S-matrix elements. For a similar discussion of how field redefintion invariant beta functions can be defined in the SMEFT, see \cite{Einhorn:2001kj}.}
The power counting expansion of the SMEFT is
relevant for gauge fixing and cannot be removed with gauge
independent field redefinitions, which is a novel feature compared to more familiar EFTs
without a Higgsed phase. 
The particular form of the field space metrics depends on the operator basis used, but the utility of the geometric approach developed here does not.
This argues for a modified gauge fixing procedure using the BFM
in the SMEFT.

{\bf Gauge fixing.}
Eliminating bilinear kinetic mixing between the gauge bosons and the Goldstone
bosons in an efficient gauge fixing procedure is advantageous. A simpler LSZ procedure \cite{Lehmann:1954rq} to
construct $S$-matrix elements results from this condition being imposed. $R_\xi$
gauge \cite{tHooft:1972qbu} in the SM when $\xi_W=\xi_B$ has some further advantages in eliminating contact operators
that complicate calculations in intermediate steps. Using the BFM combined with $R_\xi$
gauge fixing, the gauge fixing term for the $\rm SU(2)_L \times U(1)_Y$
fields in the SM takes the form
\cite{Shore:1981mj,Einhorn:1988tc,Denner:1994nn,Denner:1994xt}
\begin{align}\label{BFMSM}
	\Lagr_{\textrm{GF}}=&-\frac{1}{2\xi_W}\sum_a\left[\partial_{\mu}W^{a,\mu} -
		g_2\epsilon^{abc}\hat{W}_{b,\mu}W^{\mu}_c
		\right.	 \\
		&+ \left.ig_2\frac{\xi_W}{2}\left(\hat{H}_i^{\dagger}(\sigma^a)^i_j H^j
		- H_i^{\dagger} (\sigma^a)^i_j \hat{H}^j\right)\right]^2 \nonumber\\
		&- \frac{1}{2\xi_B}\left[ \partial_{\mu}B^{\mu} + ig_1\frac{\xi_B}{2}
		\left(\hat{H}_i^{\dagger}H^i - H_i^{\dagger}\hat{H}^i\right)\right]^2, \nonumber
\end{align}
where the background fields are denoted by a hat.

The $\rm SU(2)_L$ Pauli matrix representation in Eq.~\ref{BFMSM} is inconvenient for characterizing the gauge fixing term
as $g_{AB}$ is defined on $\mathbb{R}^4$. The Pauli matrix algebra
is isomorphic to the Clifford algebra $C(0,3)$, and the latter can be embedded in the $\mathbb{R}^4$ field
space using the real representations $\gamma_{1,2,3}$ such that
\begin{align}
	\gamma_{1,J}^{I} &=  \begin{bmatrix}
		0 & 0 & 0 & -1 \\
		0 & 0 & -1 & 0 \\
		0 & 1 & 0 & 0 \\
		1 & 0 & 0 & 0
	\end{bmatrix}, &
	\gamma_{2,J}^{I} &= \begin{bmatrix}
		0 & 0 & 1 & 0 \\
		0 & 0 & 0 & -1 \\
		-1 & 0 & 0 & 0 \\
		0 & 1 & 0 & 0
	\end{bmatrix}, \nonumber \\
	\gamma_{3,J}^{I} &=  \begin{bmatrix}
		0 & -1 & 0 & 0 \\
		1 & 0 & 0 & 0 \\
		0 & 0 & 0 & -1 \\
		0 & 0 & 1 & 0
	\end{bmatrix}, &
	\gamma_{4,J}^{I} &= \begin{bmatrix}
		0 & -1 & 0 & 0 \\
		1 & 0 & 0 & 0 \\
		0 & 0 & 0 & 1 \\
		0 & 0 & -1 & 0
	\end{bmatrix}.
\end{align}
The  $\gamma_{4}$ generator is used for the $\rm U(1)_Y$ embedding.
As $\rm SU(2)_L$ is self adjoint we can also define this algebra for the adjoint
fields, using the same real representations. $\gamma_{1,2,3,4} \langle \phi \rangle \neq 0$
and the unbroken combination of generators $(\gamma_3 + \gamma_4) \langle \phi \rangle = 0$ corresponds to $\rm U(1)_{em}$.
We absorb the couplings into the structure constants and gamma matrices,
\begin{align}
	\tilde{\epsilon}^{A}_{\, \,BC} &= g_2 \, \epsilon^{A}_{\, \, BC}, \text{ \, \, with } \tilde{\epsilon}^{1}_{\, \, 23} = +g_2,  \nonumber \\
	\tilde{\gamma}_{A,J}^{I} &= \begin{cases} g_2 \, \gamma^{I}_{A,J}, & \text{for } A=1,2,3 \\
		g_1\gamma^{I}_{A,J}, & \text{for } A=4.
					\end{cases}
\end{align}
The different couplings $g_1,g_2$ enter as the group defined on the $\mathbb{R}^4$
field space is not simple.
The $\gamma_{a,J}^{I}$ matricies satisfy the algebra $[\tilde{\gamma}_{a},\tilde{\gamma}_{b}] = 2 \, \tilde{\epsilon}_{ \, \, ab}^c \, \tilde{\gamma}_{c}$
and $[\tilde{\gamma}_{a},\tilde{\gamma}_{4}] =0$.
The structure constants vanish when any of $A,B,C=4$.
Note also that $\hat H^{\dagger}\sigma_AH-H^{\dagger}\sigma_A\hat H = -i \phi\gamma_A \hat\phi$,
with $\sigma_4 = Y =  \mathbb{I}_{2 \times 2}$.
The gauge fixing term in the background field gauge takes the form

\begin{align}\label{gaugefixing1}
	\Lagr_{\textrm{GF}} &= -\frac{\hat{g}_{AB}}{2 \, \xi} \mathcal{G}^A \, \mathcal{G}^B, \\
\mathcal{G}^X &\equiv \partial_{\mu} \mathcal{W}^{X,\mu} -
		\tilde\epsilon^{X}_{ \, \,CD}\hat{\mathcal{W}}_{\mu}^C \mathcal{W}^{D,\mu}
    + \frac{\xi}{2}\hat{g}^{XC}
		\phi^{I} \, \hat{h}_{IK} \, \tilde\gamma^{K}_{C,J} \hat{\phi}^J. \nonumber
\end{align}
The $R_{\xi}$ gauge fixing term follows when replacing the background fields with
their vacuum expectation values. The gauge fixing term is bilinear in the quantum fields.
The field space metrics in Eq.~\ref{gaugefixing1} are denoted with a hat superscript
indicating they are defined to depend only on the background fields.
Contracting with the field space metrics is a basis independent feature of the gauge fixing term.

It is useful to note the following background field gauge transformations ($\delta \hat{F}$),
with infinitesimal local gauge parameters $\delta \hat{\alpha}_A(x)$ when verifying the explicitly
the background field gauge invariance of this expression
\begin{align}\label{backgroundfieldshifts}
\delta \, \hat{\phi}^I &= -\delta \hat{\alpha}^A \, \frac{\tilde{\gamma}_{A,J}^I}{2} \hat{\phi}^J, \nonumber \\
\delta \, (D^\mu \hat{\phi})^I &= -\delta \hat{\alpha}^A \, \frac{\tilde{\gamma}_{A,J}^I}{2} (D^\mu \hat{\phi})^J, \nonumber\\
\delta \hat{\mathcal{W}}^{A, \mu} &= -\partial^\mu (\delta \hat{\alpha}^A) - \tilde{\epsilon}^A_{\, \,BC} \, \delta \hat{\alpha}^B \, \hat{\mathcal{W}}^{C, \mu}, \nonumber \\
\delta \hat{h}_{IJ} &= \hat{h}_{KJ} \, \frac{\delta \hat{\alpha}^A  \, \tilde{\gamma}_{A,I}^K}{2}+ \hat{h}_{IK} \, \frac{\delta \hat{\alpha}^A  \, \tilde{\gamma}_{A,J}^K}{2}, \nonumber \\
\delta \hat{\mathcal{W}}^A_{\mu \nu} &= - \tilde{\epsilon}^A_{\, \,BC} \, \delta \hat{\alpha}^B \, \hat{\mathcal{W}}^C_{\mu \nu}, \nonumber\\
\delta \hat{g}_{AB} &= \hat{g}_{CB} \,\tilde{\epsilon}^C_{\, \,DA} \, \delta \hat{\alpha}^D + \hat{g}_{AC} \,\tilde{\epsilon}^C_{\, \,DB} \, \delta \hat{\alpha}^D.
\end{align}
The background field gauge invariance is established by using these transformations in conjuction with
a linear change of variables on the quantum fields
\begin{align}
	\mathcal{W}^{A, \mu} &\rightarrow {\mathcal{W}}^{A, \mu} - \tilde{\epsilon}^A_{\, \,BC} \, \delta \hat{\alpha}^B \, {\mathcal{W}}^{C,\mu}, \nonumber\\
\phi^I &\rightarrow {\phi}^I - \frac{\delta \hat{\alpha}^B  \, \tilde{\gamma}_{B,K}^I}{2} {\phi}^K.
\end{align}
The transformation of the gauge fixing term is
\begin{align}
\delta \mathcal{G}^X = -\tilde{\epsilon}^X_{\, \,AB} \, \delta \hat{\alpha}^A \mathcal{G}^B.
\end{align}
With these transformations, the background field gauge invariance of the gauge fixing term is directly established.

The background field generating functional ($Z$) depends on the background fields $\hat{F} \equiv \{\hat{\mathcal{W}}^A,\hat{\phi}^I\}$
and the sources $J_F \equiv \{J^A,J^I_\phi\}$.
The source terms transform as
\begin{align}
\delta J^A_{\mu} &= - \tilde{\epsilon}^A_{\, \,BC} \, \delta \hat{\alpha}^B \, J^{C}_{\mu}, &
\delta J^I_{\phi} &= - \frac{\delta \hat{\alpha}^B  \, \tilde{\gamma}_{B,K}^I}{2} J^K_{\phi}.
\end{align}
The background field generating functional dependence on the source terms is invariant under the background field gauge transformations, as they are contracted
with the field space metrics in $Z[\hat{F},J_F]$ defined by
\begin{align}
\int \mathcal{D} F \,{\rm det}\left[\frac{\Delta \mathcal{G}^A}{\Delta \alpha^B}\right]e^{i \left(S[F + \hat{F}] + \Lagr_{\textrm{GF}} +
\hat{g}_{CD} J^C_\mu  \mathcal{W}^{D,\mu} + \hat{h}_{IJ} J^I_{\phi}  \phi^J\right)} \nonumber.
\end{align}
The integration over $d x^4$ is implicit in this expression.
Here a quantum field gauge transformation is indicated with a $\Delta$. The action is manifestly invariant
under the gauge transformation of $F+\hat{F}$.
This establishes the background field invariance of the generating functional.

The quantum fields gauge transformations are
\begin{align}
	\Delta \mathcal{W}_{\mu}^A &= -\partial_{\mu}\Delta\alpha^A - \tilde\epsilon^A_{\, \,BC} \, \Delta\alpha^B \, (\mathcal{W}_{\mu}^C
	+ \hat{\mathcal{W}}_{\mu}^C), \nonumber \\
	 \Delta \phi^I &=  - \Delta\alpha^A \, \frac{\tilde\gamma_{A,J}^{I}}{2}\, (\phi^J+ \hat{\phi}^J).
\end{align}
As the field metrics in Eq.~\ref{gaugefixing1} depend only on the background fields and do not transform under
quantum field gauge transformations, the Faddeev-Popov \cite{Faddeev:1967fc} ghost term
still follows directly; we find
\begin{align}
	\Lagr_{\textrm{FP}} = &- \hat{g}_{AB}\bar{u}^B \left[- \partial^2\delta^A_C -
		\overleftarrow\partial_{\mu}\tilde\epsilon^A_{\, \,DC}(\mathcal{W}^{D,\mu} + \hat{\mathcal{W}}^{D,\mu})\right. \nonumber\\
		&+ \tilde\epsilon^A_{\, \,DC}\hat{\mathcal{W}}^D_{\mu}\overrightarrow\partial^{\mu}
		- \tilde\epsilon^A_{\, \,DE}\tilde\epsilon^E_{\, \,FC}\hat{\mathcal{W}}^D_{\mu}
		(\mathcal{W}^{F,\mu} + \hat{\mathcal{W}}^{F,\mu}) \nonumber \\
	&- \left.\frac{\xi}{4} \hat{g}^{AD}(\phi^J + \hat{\phi}^J) \tilde\gamma_{C,J}^{I} \, \hat{h}_{IK}\,  \tilde\gamma_{D,L}^{K} \,
	\hat{\phi}^L) \right]u^C.
\end{align}
The form of this expression follows from the convention choice in Eq.~\ref{WBlagrangian}, and the descendent convention
in Eq.~\ref{gaugefixing1}.
The mass eigenstate $\mathcal{Z}_{\mu}$, $\mathcal{A}_{\mu}$ fields are defined by
\begin{align}
	\begin{bmatrix} W_{\mu}^3 \\ B_{\mu}\end{bmatrix}
	&= \begin{bmatrix} 1+ \frac{C_{HW} \bar{v}_T^2}{\Lambda^2} & -\frac{C_{HWB} \bar{v}_T^2}{2\Lambda^2} \\
		-\frac{C_{HWB} \bar{v}_T^2}{2\Lambda^2}  & 1 + \frac{C_{HB} \bar{v}_T^2}{\Lambda^2}   \end{bmatrix}
	\begin{bmatrix} c_{\bar{\theta}} & s_{\bar{\theta}} \\ -s_{\bar{\theta}} & c_{\bar{\theta}} \end{bmatrix}
	\begin{bmatrix} \mathcal{Z}_{\mu} \\ \mathcal{A}_{\mu} \end{bmatrix},\nonumber
\end{align}
where the introduced rotation angles $s_{\bar{\theta}},c_{\bar{\theta}}$ are \cite{Grinstein:1991cd,Alonso:2013hga}
\begin{align}
	t_{\bar{\theta}} \equiv \frac{s_{\bar{\theta}}}{c_{\bar{\theta}}} = \frac{\bar{g}_1}{\bar{g}_2} + \frac{\bar{v}_T^2}{2} \frac{C_{HWB}}{\Lambda^2}\left(1- \frac{\bar{g}^2_1}{\bar{g}^2_2}\right),
\end{align}
and $\bar{g}_2 = g_2 (1+ C_{HW} \bar{v}_T^2/\Lambda^2)$, $\bar{g}_1 = g_1 (1+ C_{HB} \bar{v}_T^2/\Lambda^2)$.
This removes mixing terms as well as making the kinetic term of the spin one electroweak fields canonically normalized.
This results in a simplified LSZ procedure to construct S-matrix elements.
Ghost fields associated with the mass eigenstates follow from the linear rotation to the mass eigenstate fields.
Feynman rules can be extracted directly from these expressions. Corrections from the
higher dimensional operators ($C_{H\Box},C_{HD},C_{HWB},C_{HB},C_{HW}$) enter in ghost interactions and couple to the sources
through the gauge and scalar metrics.

{\bf Conclusions}
In this paper we have defined an approach to gauge fixing the SMEFT that preserves background field gauge invariance.
This approach directly generalizes to higher orders in the SMEFT power counting. The key point is to gauge fix the fields on the curved
field space due to the power counting expansion.

%\begin{align}
%	\Lagr_{\textrm{FP}} = - g_{AB}\bar{u}^B\left[ \partial^2\delta^A_C -
%		\overleftarrow\partial_{\mu}\tilde\epsilon^A_{DC}(W^{D,\mu} + \hat{W}^{D,\mu})\right. \nonumber\\
%		+ \tilde\epsilon^A_{DC}\hat{W}^D_{\mu}\overrightarrow\partial^{\mu}
%		+ \tilde\epsilon^A_{DE}\tilde\epsilon^E_{FC}\hat{W}^D_{\mu}
%		(W^{F,\mu} + \hat{W}^{F,\mu}) \nonumber \\
%	+ \left.i\frac{\xi}{2}g^{AD}h_{IK}h_{JL}(\hat\phi^I\tilde\gamma_D^{KL}
%	\tilde\gamma_C^{JM}\phi_M) \left]u^C.
%\end{align}

%\begin{acknowledgments}
{\bf Acknowledgements}
MT and AH acknowledge support from the Villum Fonden and   the Danish National Research Foundation (DNRF91)
  through the Discovery center. MT is  grateful to the Mainz Institute for Theoretical Physics (MITP)
  for hospitality and partial support during the completion of this work.
We thank E. Bjerrum-Bohr, I. Brivio, P. Damgaard, C. Hartmann,
  A. Manohar, and G. Passarino for useful discussions related to this material and/or comments on the draft.
%\end{acknowledgments}


\bibliography{bibliography_V2}% Produces the bibliography via BibTeX.

\begin{thebibliography}{24}%
\makeatletter
\providecommand \@ifxundefined [1]{%
 \@ifx{#1\undefined}
}%
\providecommand \@ifnum [1]{%
 \ifnum #1\expandafter \@firstoftwo
 \else \expandafter \@secondoftwo
 \fi
}%
\providecommand \@ifx [1]{%
 \ifx #1\expandafter \@firstoftwo
 \else \expandafter \@secondoftwo
 \fi
}%
\providecommand \natexlab [1]{#1}%
\providecommand \enquote  [1]{``#1''}%
\providecommand \bibnamefont  [1]{#1}%
\providecommand \bibfnamefont [1]{#1}%
\providecommand \citenamefont [1]{#1}%
\providecommand \href@noop [0]{\@secondoftwo}%
\providecommand \href [0]{\begingroup \@sanitize@url \@href}%
\providecommand \@href[1]{\@@startlink{#1}\@@href}%
\providecommand \@@href[1]{\endgroup#1\@@endlink}%
\providecommand \@sanitize@url [0]{\catcode `\\12\catcode `\$12\catcode
  `\&12\catcode `\#12\catcode `\^12\catcode `\_12\catcode `\%12\relax}%
\providecommand \@@startlink[1]{}%
\providecommand \@@endlink[0]{}%
\providecommand \url  [0]{\begingroup\@sanitize@url \@url }%
\providecommand \@url [1]{\endgroup\@href {#1}{\urlprefix }}%
\providecommand \urlprefix  [0]{URL }%
\providecommand \Eprint [0]{\href }%
\providecommand \doibase [0]{http://dx.doi.org/}%
\providecommand \selectlanguage [0]{\@gobble}%
\providecommand \bibinfo  [0]{\@secondoftwo}%
\providecommand \bibfield  [0]{\@secondoftwo}%
\providecommand \translation [1]{[#1]}%
\providecommand \BibitemOpen [0]{}%
\providecommand \bibitemStop [0]{}%
\providecommand \bibitemNoStop [0]{.\EOS\space}%
\providecommand \EOS [0]{\spacefactor3000\relax}%
\providecommand \BibitemShut  [1]{\csname bibitem#1\endcsname}%
\let\auto@bib@innerbib\@empty
%</preamble>
\bibitem [{\citenamefont {Alonso}\ \emph
  {et~al.}(2016{\natexlab{a}})\citenamefont {Alonso}, \citenamefont {Jenkins},\
  and\ \citenamefont {Manohar}}]{Alonso:2015fsp}%
  \BibitemOpen
  \bibfield  {author} {\bibinfo {author} {\bibfnamefont {R.}~\bibnamefont
  {Alonso}}, \bibinfo {author} {\bibfnamefont {E.~E.}\ \bibnamefont {Jenkins}},
  \ and\ \bibinfo {author} {\bibfnamefont {A.~V.}\ \bibnamefont {Manohar}},\
  }\href {\doibase 10.1016/j.physletb.2016.01.041} {\bibfield  {journal}
  {\bibinfo  {journal} {Phys. Lett.}\ }\textbf {\bibinfo {volume} {B754}},\
  \bibinfo {pages} {335} (\bibinfo {year} {2016}{\natexlab{a}})},\ \Eprint
  {http://arxiv.org/abs/1511.00724} {arXiv:1511.00724} \BibitemShut {NoStop}%
%%CITATION = ARXIV:1511.00724;%%
\bibitem [{\citenamefont {Alonso}\ \emph
  {et~al.}(2016{\natexlab{b}})\citenamefont {Alonso}, \citenamefont {Jenkins},\
  and\ \citenamefont {Manohar}}]{Alonso:2016oah}%
  \BibitemOpen
  \bibfield  {author} {\bibinfo {author} {\bibfnamefont {R.}~\bibnamefont
  {Alonso}}, \bibinfo {author} {\bibfnamefont {E.~E.}\ \bibnamefont {Jenkins}},
  \ and\ \bibinfo {author} {\bibfnamefont {A.~V.}\ \bibnamefont {Manohar}},\
  }\href {\doibase 10.1007/JHEP08(2016)101} {\bibfield  {journal} {\bibinfo
  {journal} {JHEP}\ }\textbf {\bibinfo {volume} {08}},\ \bibinfo {pages} {101}
  (\bibinfo {year} {2016}{\natexlab{b}})},\ \Eprint
  {http://arxiv.org/abs/1605.03602} {arXiv:1605.03602} \BibitemShut {NoStop}%
%%CITATION = ARXIV:1605.03602;%%
\bibitem [{\citenamefont {DeWitt}(1967)}]{DeWitt:1967ub}%
  \BibitemOpen
  \bibfield  {author} {\bibinfo {author} {\bibfnamefont {B.~S.}\ \bibnamefont
  {DeWitt}},\ }\href {\doibase 10.1103/PhysRev.162.1195} {\bibfield  {journal}
  {\bibinfo  {journal} {Phys. Rev.}\ }\textbf {\bibinfo {volume} {162}},\
  \bibinfo {pages} {1195} (\bibinfo {year} {1967})}\BibitemShut {NoStop}%
%%CITATION = PHRVA,162,1195;%%
\bibitem [{\citenamefont {'t~Hooft}(1973)}]{tHooft:1973bhk}%
  \BibitemOpen
  \bibfield  {author} {\bibinfo {author} {\bibfnamefont {G.}~\bibnamefont
  {'t~Hooft}},\ }\href {\doibase 10.1016/0550-3213(73)90263-0} {\bibfield
  {journal} {\bibinfo  {journal} {Nucl. Phys.}\ }\textbf {\bibinfo {volume}
  {B62}},\ \bibinfo {pages} {444} (\bibinfo {year} {1973})}\BibitemShut
  {NoStop}%
%%CITATION = NUPHA,B62,444;%%
\bibitem [{\citenamefont {Abbott}(1982)}]{Abbott:1981ke}%
  \BibitemOpen
  \bibfield  {author} {\bibinfo {author} {\bibfnamefont {L.~F.}\ \bibnamefont
  {Abbott}},\ }\href@noop {} {\bibfield  {journal} {\bibinfo  {journal} {Acta
  Phys. Polon.}\ }\textbf {\bibinfo {volume} {B13}},\ \bibinfo {pages} {33}
  (\bibinfo {year} {1982})}\BibitemShut {NoStop}%
%%CITATION = APPOA,B13,33;%%
\bibitem [{\citenamefont {Shore}(1981)}]{Shore:1981mj}%
  \BibitemOpen
  \bibfield  {author} {\bibinfo {author} {\bibfnamefont {G.~M.}\ \bibnamefont
  {Shore}},\ }\href {\doibase 10.1016/0003-4916(81)90198-6} {\bibfield
  {journal} {\bibinfo  {journal} {Annals Phys.}\ }\textbf {\bibinfo {volume}
  {137}},\ \bibinfo {pages} {262} (\bibinfo {year} {1981})}\BibitemShut
  {NoStop}%
%%CITATION = APNYA,137,262;%%
\bibitem [{\citenamefont {Einhorn}\ and\ \citenamefont
  {Wudka}(1989)}]{Einhorn:1988tc}%
  \BibitemOpen
  \bibfield  {author} {\bibinfo {author} {\bibfnamefont {M.}~\bibnamefont
  {Einhorn}}\ and\ \bibinfo {author} {\bibfnamefont {J.}~\bibnamefont
  {Wudka}},\ }\href {\doibase 10.1103/PhysRevD.39.2758} {\bibfield  {journal}
  {\bibinfo  {journal} {Phys. Rev.}\ }\textbf {\bibinfo {volume} {D39}},\
  \bibinfo {pages} {2758} (\bibinfo {year} {1989})}\BibitemShut {NoStop}%
%%CITATION = PHRVA,D39,2758;%%
\bibitem [{\citenamefont {Denner}\ \emph {et~al.}(1994)\citenamefont {Denner},
  \citenamefont {Weiglein},\ and\ \citenamefont {Dittmaier}}]{Denner:1994nn}%
  \BibitemOpen
  \bibfield  {author} {\bibinfo {author} {\bibfnamefont {A.}~\bibnamefont
  {Denner}}, \bibinfo {author} {\bibfnamefont {G.}~\bibnamefont {Weiglein}}, \
  and\ \bibinfo {author} {\bibfnamefont {S.}~\bibnamefont {Dittmaier}},\ }\href
  {\doibase 10.1016/0370-2693(94)90162-7} {\bibfield  {journal} {\bibinfo
  {journal} {Phys. Lett.}\ }\textbf {\bibinfo {volume} {B333}},\ \bibinfo
  {pages} {420} (\bibinfo {year} {1994})},\ \Eprint
  {http://arxiv.org/abs/hep-ph/9406204} {arXiv:hep-ph/9406204} \BibitemShut
  {NoStop}%
%%CITATION = HEP-PH/9406204;%%
\bibitem [{\citenamefont {Denner}\ \emph {et~al.}(1995)\citenamefont {Denner},
  \citenamefont {Weiglein},\ and\ \citenamefont {Dittmaier}}]{Denner:1994xt}%
  \BibitemOpen
  \bibfield  {author} {\bibinfo {author} {\bibfnamefont {A.}~\bibnamefont
  {Denner}}, \bibinfo {author} {\bibfnamefont {G.}~\bibnamefont {Weiglein}}, \
  and\ \bibinfo {author} {\bibfnamefont {S.}~\bibnamefont {Dittmaier}},\ }\href
  {\doibase 10.1016/0550-3213(95)00037-S} {\bibfield  {journal} {\bibinfo
  {journal} {Nucl. Phys.}\ }\textbf {\bibinfo {volume} {B440}},\ \bibinfo
  {pages} {95} (\bibinfo {year} {1995})},\ \Eprint
  {http://arxiv.org/abs/hep-ph/9410338} {arXiv:hep-ph/9410338 [hep-ph]}
  \BibitemShut {NoStop}%
%%CITATION = HEP-PH/9410338;%%
\bibitem [{\citenamefont {Hartmann}\ and\ \citenamefont
  {Trott}(2015)}]{Hartmann:2015oia}%
  \BibitemOpen
  \bibfield  {author} {\bibinfo {author} {\bibfnamefont {C.}~\bibnamefont
  {Hartmann}}\ and\ \bibinfo {author} {\bibfnamefont {M.}~\bibnamefont
  {Trott}},\ }\href {\doibase 10.1007/JHEP07(2015)151} {\bibfield  {journal}
  {\bibinfo  {journal} {JHEP}\ }\textbf {\bibinfo {volume} {07}},\ \bibinfo
  {pages} {151} (\bibinfo {year} {2015})},\ \Eprint
  {http://arxiv.org/abs/1505.02646} {arXiv:1505.02646} \BibitemShut {NoStop}%
%%CITATION = ARXIV:1505.02646;%%
\bibitem [{\citenamefont {Ward}(1950)}]{Ward:1950xp}%
  \BibitemOpen
  \bibfield  {author} {\bibinfo {author} {\bibfnamefont {J.~C.}\ \bibnamefont
  {Ward}},\ }\href {\doibase 10.1103/PhysRev.78.182} {\bibfield  {journal}
  {\bibinfo  {journal} {Phys. Rev.}\ }\textbf {\bibinfo {volume} {78}},\
  \bibinfo {pages} {182} (\bibinfo {year} {1950})}\BibitemShut {NoStop}%
%%CITATION = PHRVA,78,182;%%
\bibitem [{\citenamefont {Becchi}\ \emph {et~al.}(1976)\citenamefont {Becchi},
  \citenamefont {Rouet},\ and\ \citenamefont {Stora}}]{Becchi:1975nq}%
  \BibitemOpen
  \bibfield  {author} {\bibinfo {author} {\bibfnamefont {C.}~\bibnamefont
  {Becchi}}, \bibinfo {author} {\bibfnamefont {A.}~\bibnamefont {Rouet}}, \
  and\ \bibinfo {author} {\bibfnamefont {R.}~\bibnamefont {Stora}},\ }\href
  {\doibase 10.1016/0003-4916(76)90156-1} {\bibfield  {journal} {\bibinfo
  {journal} {Annals Phys.}\ }\textbf {\bibinfo {volume} {98}},\ \bibinfo
  {pages} {287} (\bibinfo {year} {1976})}\BibitemShut {NoStop}%
%%CITATION = APNYA,98,287;%%
\bibitem [{\citenamefont {Dedes}\ \emph {et~al.}(2017)\citenamefont {Dedes},
  \citenamefont {Materkowska}, \citenamefont {Paraskevas}, \citenamefont
  {Rosiek},\ and\ \citenamefont {Suxho}}]{Dedes:2017zog}%
  \BibitemOpen
  \bibfield  {author} {\bibinfo {author} {\bibfnamefont {A.}~\bibnamefont
  {Dedes}}, \bibinfo {author} {\bibfnamefont {W.}~\bibnamefont {Materkowska}},
  \bibinfo {author} {\bibfnamefont {M.}~\bibnamefont {Paraskevas}}, \bibinfo
  {author} {\bibfnamefont {J.}~\bibnamefont {Rosiek}}, \ and\ \bibinfo {author}
  {\bibfnamefont {K.}~\bibnamefont {Suxho}},\ }\href {\doibase
  10.1007/JHEP06(2017)143} {\bibfield  {journal} {\bibinfo  {journal} {JHEP}\
  }\textbf {\bibinfo {volume} {06}},\ \bibinfo {pages} {143} (\bibinfo {year}
  {2017})},\ \Eprint {http://arxiv.org/abs/1704.03888} {arXiv:1704.03888}
  \BibitemShut {NoStop}%
%%CITATION = ARXIV:1704.03888;%%
\bibitem [{\citenamefont {Grzadkowski}\ \emph {et~al.}(2010)\citenamefont
  {Grzadkowski}, \citenamefont {Iskrzynski}, \citenamefont {Misiak},\ and\
  \citenamefont {Rosiek}}]{Grzadkowski:2010es}%
  \BibitemOpen
  \bibfield  {author} {\bibinfo {author} {\bibfnamefont {B.}~\bibnamefont
  {Grzadkowski}}, \bibinfo {author} {\bibfnamefont {M.}~\bibnamefont
  {Iskrzynski}}, \bibinfo {author} {\bibfnamefont {M.}~\bibnamefont {Misiak}},
  \ and\ \bibinfo {author} {\bibfnamefont {J.}~\bibnamefont {Rosiek}},\ }\href
  {\doibase 10.1007/JHEP10(2010)085} {\bibfield  {journal} {\bibinfo  {journal}
  {JHEP}\ }\textbf {\bibinfo {volume} {10}},\ \bibinfo {pages} {085} (\bibinfo
  {year} {2010})},\ \Eprint {http://arxiv.org/abs/1008.4884} {arXiv:1008.4884
  [hep-ph]} \BibitemShut {NoStop}%
%%CITATION = ARXIV:1008.4884;%%
\bibitem [{\citenamefont {Burgess}\ \emph {et~al.}(2010)\citenamefont
  {Burgess}, \citenamefont {Lee},\ and\ \citenamefont
  {Trott}}]{Burgess:2010zq}%
  \BibitemOpen
  \bibfield  {author} {\bibinfo {author} {\bibfnamefont {C.~P.}\ \bibnamefont
  {Burgess}}, \bibinfo {author} {\bibfnamefont {H.~M.}\ \bibnamefont {Lee}}, \
  and\ \bibinfo {author} {\bibfnamefont {M.}~\bibnamefont {Trott}},\ }\href
  {\doibase 10.1007/JHEP07(2010)007} {\bibfield  {journal} {\bibinfo  {journal}
  {JHEP}\ }\textbf {\bibinfo {volume} {07}},\ \bibinfo {pages} {007} (\bibinfo
  {year} {2010})},\ \Eprint {http://arxiv.org/abs/1002.2730} {arXiv:1002.2730}
  \BibitemShut {NoStop}%
%%CITATION = ARXIV:1002.2730;%%
\bibitem [{\citenamefont {Passarino}\ and\ \citenamefont
  {Trott}(2016)}]{Passarino:2016pzb}%
  \BibitemOpen
  \bibfield  {author} {\bibinfo {author} {\bibfnamefont {G.}~\bibnamefont
  {Passarino}}\ and\ \bibinfo {author} {\bibfnamefont {M.}~\bibnamefont
  {Trott}},\ }\href@noop {} {\  (\bibinfo {year} {2016})},\ \Eprint
  {http://arxiv.org/abs/1610.08356} {arXiv:1610.08356} \BibitemShut {NoStop}%
%%CITATION = ARXIV:1610.08356;%%
\bibitem [{\citenamefont {Passarino}(2017)}]{Passarino:2016saj}%
  \BibitemOpen
  \bibfield  {author} {\bibinfo {author} {\bibfnamefont {G.}~\bibnamefont
  {Passarino}},\ }\href {\doibase 10.1140/epjp/i2017-11291-5} {\bibfield
  {journal} {\bibinfo  {journal} {Eur. Phys. J. Plus}\ }\textbf {\bibinfo
  {volume} {132}},\ \bibinfo {pages} {16} (\bibinfo {year} {2017})},\ \Eprint
  {http://arxiv.org/abs/1610.09618} {arXiv:1610.09618} \BibitemShut {NoStop}%
%%CITATION = ARXIV:1610.09618;%%
\bibitem [{\citenamefont {Brivio}\ and\ \citenamefont
  {Trott}(2017)}]{Brivio:2017vri}%
  \BibitemOpen
  \bibfield  {author} {\bibinfo {author} {\bibfnamefont {I.}~\bibnamefont
  {Brivio}}\ and\ \bibinfo {author} {\bibfnamefont {M.}~\bibnamefont {Trott}},\
  }\href@noop {} {\  (\bibinfo {year} {2017})},\ \Eprint
  {http://arxiv.org/abs/1706.08945} {arXiv:1706.08945} \BibitemShut {NoStop}%
%%CITATION = ARXIV:1706.08945;%%
\bibitem [{\citenamefont {Einhorn}\ and\ \citenamefont
  {Wudka}(2001)}]{Einhorn:2001kj}%
  \BibitemOpen
  \bibfield  {author} {\bibinfo {author} {\bibfnamefont {M.~B.}\ \bibnamefont
  {Einhorn}}\ and\ \bibinfo {author} {\bibfnamefont {J.}~\bibnamefont
  {Wudka}},\ }\href {\doibase 10.1088/1126-6708/2001/08/025} {\bibfield
  {journal} {\bibinfo  {journal} {JHEP}\ }\textbf {\bibinfo {volume} {08}},\
  \bibinfo {pages} {025} (\bibinfo {year} {2001})},\ \Eprint
  {http://arxiv.org/abs/hep-ph/0105035} {arXiv:hep-ph/0105035 [hep-ph]}
  \BibitemShut {NoStop}%
%%CITATION = HEP-PH/0105035;%%
\bibitem [{\citenamefont {Lehmann}\ \emph {et~al.}(1955)\citenamefont
  {Lehmann}, \citenamefont {Symanzik},\ and\ \citenamefont
  {Zimmermann}}]{Lehmann:1954rq}%
  \BibitemOpen
  \bibfield  {author} {\bibinfo {author} {\bibfnamefont {H.}~\bibnamefont
  {Lehmann}}, \bibinfo {author} {\bibfnamefont {K.}~\bibnamefont {Symanzik}}, \
  and\ \bibinfo {author} {\bibfnamefont {W.}~\bibnamefont {Zimmermann}},\
  }\href {\doibase 10.1007/BF02731765} {\bibfield  {journal} {\bibinfo
  {journal} {Nuovo Cim.}\ }\textbf {\bibinfo {volume} {1}},\ \bibinfo {pages}
  {205} (\bibinfo {year} {1955})}\BibitemShut {NoStop}%
%%CITATION = NUCIA,1,205;%%
\bibitem [{\citenamefont {'t~Hooft}\ and\ \citenamefont
  {Veltman}(1972)}]{tHooft:1972qbu}%
  \BibitemOpen
  \bibfield  {author} {\bibinfo {author} {\bibnamefont {'t~Hooft}}\ and\
  \bibinfo {author} {\bibnamefont {Veltman}},\ }\href {\doibase
  10.1016/S0550-3213(72)80021-X} {\bibfield  {journal} {\bibinfo  {journal}
  {Nucl. Phys.}\ }\textbf {\bibinfo {volume} {B50}},\ \bibinfo {pages} {318}
  (\bibinfo {year} {1972})}\BibitemShut {NoStop}%
%%CITATION = NUPHA,B50,318;%%
\bibitem [{\citenamefont {Faddeev}\ and\ \citenamefont
  {Popov}(1967)}]{Faddeev:1967fc}%
  \BibitemOpen
  \bibfield  {author} {\bibinfo {author} {\bibfnamefont {L.}~\bibnamefont
  {Faddeev}}\ and\ \bibinfo {author} {\bibfnamefont {V.}~\bibnamefont
  {Popov}},\ }\href {\doibase 10.1016/0370-2693(67)90067-6} {\bibfield
  {journal} {\bibinfo  {journal} {Phys. Lett.}\ }\textbf {\bibinfo {volume}
  {25B}},\ \bibinfo {pages} {29} (\bibinfo {year} {1967})}\BibitemShut
  {NoStop}%
%%CITATION = PHLTA,25B,29;%%
\bibitem [{\citenamefont {Grinstein}\ and\ \citenamefont
  {Wise}(1991)}]{Grinstein:1991cd}%
  \BibitemOpen
  \bibfield  {author} {\bibinfo {author} {\bibfnamefont {B.}~\bibnamefont
  {Grinstein}}\ and\ \bibinfo {author} {\bibfnamefont {M.}~\bibnamefont
  {Wise}},\ }\href {\doibase 10.1016/0370-2693(91)90061-T} {\bibfield
  {journal} {\bibinfo  {journal} {Phys.Lett.}\ }\textbf {\bibinfo {volume}
  {B265}},\ \bibinfo {pages} {326} (\bibinfo {year} {1991})}\BibitemShut
  {NoStop}%
%%CITATION = PHLTA,B265,326;%%
\bibitem [{\citenamefont {Alonso}\ \emph {et~al.}(2014)\citenamefont {Alonso},
  \citenamefont {Jenkins}, \citenamefont {Manohar},\ and\ \citenamefont
  {Trott}}]{Alonso:2013hga}%
  \BibitemOpen
  \bibfield  {author} {\bibinfo {author} {\bibfnamefont {R.}~\bibnamefont
  {Alonso}}, \bibinfo {author} {\bibfnamefont {E.~E.}\ \bibnamefont {Jenkins}},
  \bibinfo {author} {\bibfnamefont {A.~V.}\ \bibnamefont {Manohar}}, \ and\
  \bibinfo {author} {\bibfnamefont {M.}~\bibnamefont {Trott}},\ }\href
  {\doibase 10.1007/JHEP04(2014)159} {\bibfield  {journal} {\bibinfo  {journal}
  {JHEP}\ }\textbf {\bibinfo {volume} {1404}},\ \bibinfo {pages} {159}
  (\bibinfo {year} {2014})},\ \Eprint {http://arxiv.org/abs/1312.2014}
  {arXiv:1312.2014} \BibitemShut {NoStop}%
%%CITATION = ARXIV:1312.2014;%%
\end{thebibliography}%

\end{document}